\documentclass[12pt]
{iopart}
\newcommand{\gguide}{{\it Preparing graphics for IOP journals}}
\usepackage{iopams} 
\usepackage{setstack}
\usepackage{graphicx}
\begin{document}

\title[A model for simulating speckle-pattern evolution  in OCT]{A model for simulating speckle-pattern evolution based on close to reality procedures used in spectral-domain OCT}

\author{V.Yu.~Zaitsev$^{1,3}$, L.A.~Matveev$^{1,2}$, 
A.L.~Matveyev$^{1,2}$, G.V.~Gelikonov$^{1,2}$,  and V.M. Gelikonov$^{1,3}$ }

\address{$^1$Institute of Applied Physics RAS, Uljanova St. 46, Nizhny Novgorod,  Russia}
\address{$^2$Nizhny Novgorod State Medical Academy, Minin Square 10/1 , Nizhny Novgorod,  Russia}
\address{$^3$Nizhny Novgorod State University, Gagarina Av. 23, Nizhny Novgorod,  Russia}
\ead{vyuzai@hydro.appl-sci.nnov.ru}
\begin{abstract}
A robust model for simulating speckle pattern evolution in optical coherence tomography (OCT) depending on the OCT system parameters and tissue deformation is reported. The model is based on application of close to reality procedures used in spectral-domain OCT scanners. It naturally generates images reproducing properties of real images in spectral-domain OCT, including the pixelized structure and finite depth of unambiguous imaging, influence of the  optical spectrum shape, dependence on the optical wave frequency and coherence length, influence of the tissue straining, etc. Good agreement with generally accepted speckle features and properties of real OCT images is demonstrated.

\end{abstract}

\submitto{Laser Physics Letters}
\maketitle

\section{Introduction}
The problem of modeling speckle patterns of OCT images attracts a significant attention because for many applications based on analysis of OCT images the speckle structure plays a key role. Since the first works (e.g. \cite{ref1}) related to studies of speckle patterns in OCT, a number of methods that can be used for simulating speckle patterns and studying correlation properties have been discussed. In ref.~\cite{ref1}, the notion of signal-carrying speckles and signal-degrading speckles were introduced. The former reflect the relation of speckles in OCT images with the local density of scatterers and the latter correspond to intensity variations induced by the conditions of light propagation that are not directly related to the single scattering form the localized scatterers. The methods developed for simulation of initial speckle patterns and their evolution comprise, in particular, representation of speckle amplitudes as a sum of sinusoidal functions with random phases (following [1]), application of certain functional transformations to the initial random speckle pattern in order to obtain desired variations in the speckle-correlation properties of various types (i.e. with some prescribed character of correlation evolution including both monotonic decrease and non-monotonic behavior of cross-correlation \cite{ref2}, as well as various modifications of speckle-pattern simulation based on constructing speckle spots using an assumed point spread functions (PSF) (e.g., \cite{ref3}). Besides, a popular trend is the application of Monte-Carlo approaches for taking into account phase (and amplitude) variability of waves scattered by localized scatterers. In terms of ref.~\cite{ref1}, this makes it possible to simulate both signal-carrying and signal-degrading speckles \cite{ref4,ref5}. A vast amount of works are devoted to development of speckle-reduction methods (e.g., based on compounding images obtained for waves with different frequencies \cite{ref6}, polarizations \cite{ref7}, angular orientations \cite{ref8} and various (often rather sophisticated) filtering methods \cite{ref9,ref10}. 

The development of the model reported here was stimulated first of all by the necessity of studying the evolution of speckle patterns, including correlation properties for different parameters of the OCT system, which is a key issue in the context of development of new variants of elastographic imaging in OCT \cite{ref11,ref12,ref13,ref14,ref15,ref16,ref17}. The same understanding is also indispensable for a wide range of problems related to the development of super-resolution methods for application to OCT image processing, perfection of speckle-suppression methods, etc. To study these problems, one requires a model for realistic description of speckle-pattern evolution caused by deformation of the tissue. Other important requirements to such a model include the possibility of studying the role of such characteristics as the coherence length of the optical field, the relation between the number of spectral components used for image formation in spectral-domain OCT scanners and the number of discrete elements if real pixelized images, the density of scatterers distributed in the tissue, etc. Another desired requirement is the possibility to directly obtain pixelized images with the structure as close as possible to real OCT images, so that the developed for simulated images processing could be directly applied to real images. Although the reported model is based on the procedures close to those used in spectral-domain OCT, the obtained results can actually be applied to images obtained by time-domain OCT scanners (if the appropriate ratio between the optical wave length and the coherence length is used).

 \section{Basic idea of the model}  
The realistic character of speckle-pattern simulation is ensured by the fact that the considered model reproduces (omitting technical details) main features of image-formation in real spectral-domain OCT devices. Although principles of spectral-domain OCT scanners are widely discussed in literature (see, e.g.,~
\cite{ref18,ref19,ref20,ref21,ref22,ref23} and literature cited therein), we first recall some basic issues in order to clearer emphasize key physical points of the reported model. 

Consider first the volume of the tissue limited in the lateral direction by the transversal size of the optical beam. In the axial direction, this region from one side is limited by the tissue surface $z=0$ that is convenient to associate with the middle of the first pixel in the image to be constructed. Let us assume that the other boundary in the axial direction corresponds to the middle of the last pixel with coordinate $z=H$. Here, $H$ is the desired maximal interrogation depth (although physically the tissue may continue for $z>H$). Thus a region with the axial size $0<z\le H$ corresponds to one A-scan in the OCT image, where $N$ pixels correspond to $N$ physical cells in the tissue containing scatterers as shown in Fig.\,1. 

\begin{figure}[htb]
\centerline{ \includegraphics[width=12cm]{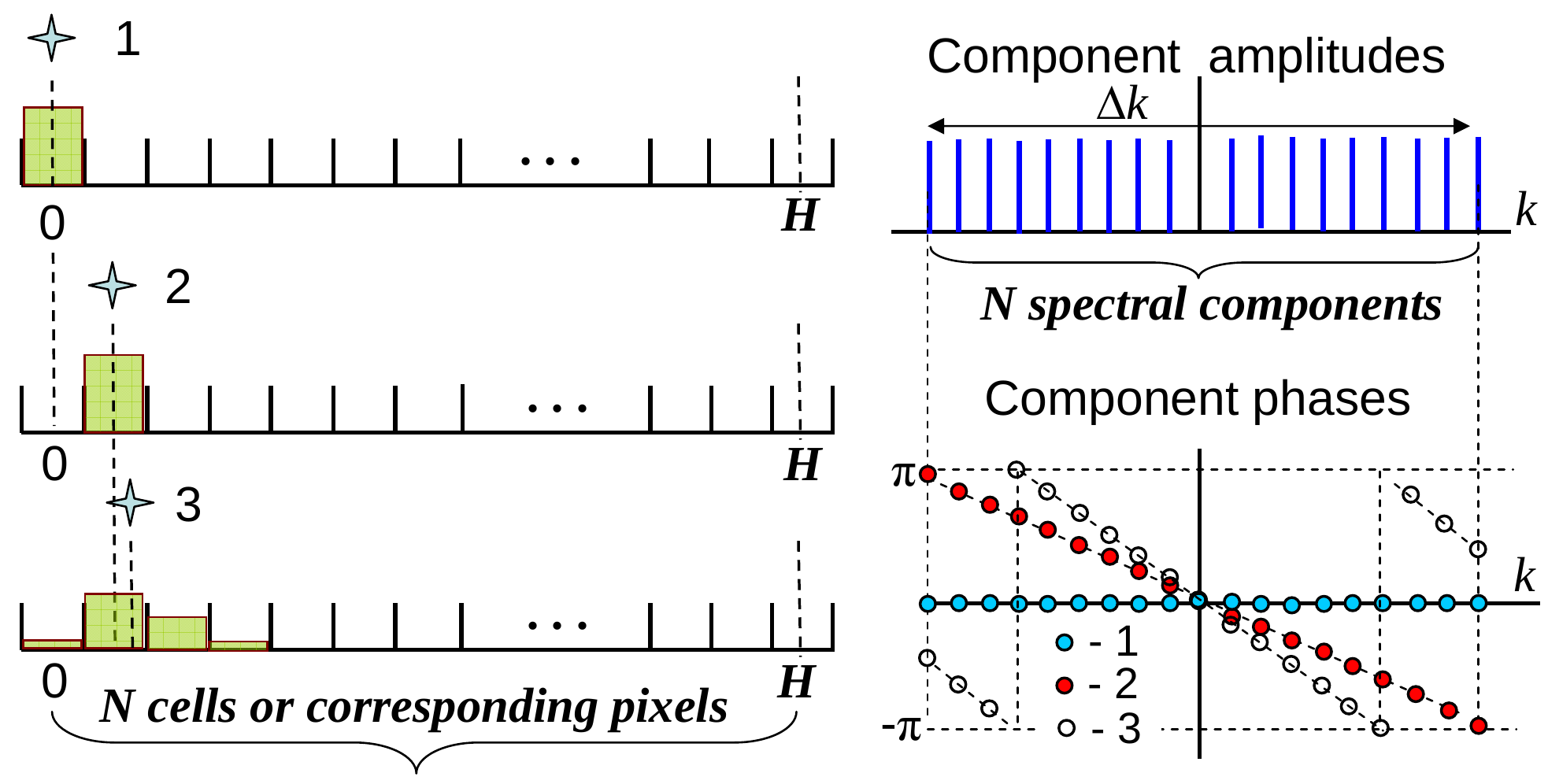} }
\caption{Schematically shown 1D array of $N$ spatial cells equivalent to pixels in an A-scan (left panels) and the corresponding Fourier components in the spatial spectrum (right panels) for the three positions $z_{0} $ of a single localized scatterer (shown by asterisk): in the center of the fist cell (at the surface $z_{0} =0$), in the center of the next cell (i.e., exactly 1 pixel deeper), and 1.25 pixel deeper. The phases of the spectral components acquire phase shifts described by the factor $\exp (-ik_{n} z_{0} )$, whereas their absolute values do not change in the absence of interference of contr0butions of several neighboring scatterers. Summation of all $N$ spectral harmonics ensures the localization of the pixelized scatterer image (shown by rectangles) within a single pixel only if the particle is exactly in the center of the cell (cases 1 and 2). For locations of a point-like scatterer between the cells' centers, its image is spread among a few neighboring pixels as schematically shown for case 3.
}
\end{figure}

Let us first assume that a unit-amplitude single element (point-like particle) is located in the first cell and has the axial coordinate $z_{0} =0$, and in all other cells the are no elements (or, equivalently, all other $(N-1)$ cells correspond to zero-amplitude elements). This array of discrete elements has a spatial Fourier spectrum such that $N$ elements in the space domain correspond to $N$ spectral harmonics $\propto \exp (\pm ik_{n} z)$ with certain complex amplitudes in the formal spatial spectrum as shown in Fig. 1 for the case of a single element. The interval $\delta k$ between the spectral harmonics is determined by the entire axial size $0\le z\le H$ of the array of the cells containing particles, $\delta k=2\pi /H$. For the element located at $z_{0} =0$, the harmonics in the spectrum $S(k_{n} ,z_{0} =0)$ have equal amplitudes and zero phases (case 1 in Fig.~1). The inverse Fourier transform corresponds to summation of these discrete spectral harmonics $S(k_{i} ,z_{0} =0)$ and yields exactly an array of zeros in all spatial cells (pixels) except for the first cell containing a single unit-amplitude scatterer with $z_{0} =0$ (case 1 in Fig.~1). For any other position of such a single scatterer, the amplitudes of the harmonics remain the same, whereas their phases are modified. Namely, for $z_{0} \ne 0$, according to the properties of Fourier-transforms, the spectral components $\exp (ik_{n} z)$ acquire phase shifts corresponding to the appearance of an additional factor $\exp (-ik_{n} z_{0} )$ for each harmonic (see cases 2 and 3 in Fig.~1). Summation of such spectral harmonics with the so-shifted phases (which is equivalent to the procedure of inverse Fourier transform) makes it possible to construct the corresponding spatial domain pattern for an arbitrary axial position $z=z_{0} $ of such a single unit-amplitude element. An important for the present discussion consequence of this procedure is that if the scatterer coordinate does not exactly coincide with the center of a cell ($z_{0} \ne nH/(N-1)$, where $n=0,\; 1,\; 2...N-1$), the space-domain pattern in an A-scan still can be correctly reconstructed. For example, if a single point-like scatterer is located between centers of the cells (case 3 in Fig.~1), the corresponding maximum in the resultant amplitude/intensity pattern becomes spread over these adjacent cells and probably a few more distant neighboring cells 

Next, if there is a set of such points/elements with arbitrarily distributed coordinates $z_{j} $ and amplitudes $A_{j} $, the entire spectrum is given by summation of individual spectra for all localized elements. Correspondingly, the interference (linear superposition) of all those spectral harmonics results in a certain spatial pattern, the intensity of which is determined by mutual phases and amplitudes of all spectral harmonics. The resultant amplitudes in $q$th pixel with  the center located at the depth distribution $z_{q} =qH/(N-1)$ (where $q=1..N-1)$ within a single A-scan can be written as 
\begin{equation} \label{GrindEQ__1_}
A(q)=\sum _{j}\sum _{n}^{} A_{j} \exp (ik_{n} z_{j} )\exp (-ik_{n} z_{q} )  
\end{equation} 
Here, for symmetry, it is assumed an odd number $N$, the index $n$ relates to summation over all spectral harmonics in the range $\mp (N-1)/2$, for which $\delta k=k_{n+1} -k_{n} =2\pi /H$, $k_{n} =2\pi n/H$, and index $j$ corresponds to summation of contributions from all particles localized within the tissue volume represented by the given A-scan. Taking into account the above-introduced notations, Eq.~(\ref{GrindEQ__1_})  can be rewritten as
\begin{equation} \label{GrindEQ__2_}
A(q)=\sum _{j}\sum _{n}^{} A_{j} \exp (i\frac{2\pi n}{H} z_{j} )\exp (-i\frac{2\pi n}{H} z_{q} )  
\end{equation} 
We recall that although the position $z_{j} $ of $j$th scatterer can be arbitrary, the axial coordinate ($z_{q} $) in the pixelized A-scan is discrete with a spatial step $H/(N-1)$ between the adjacent pixels. Correspondingly, although the different positions of a scatterer inside the $q$th pixel has an uncertainly $\pm H/2N$, the amplitude distribution for several neighboring pixels should be somewhat different depending on the exact coordinate of the element close to $qH/(N-1)$. 

To physically implement in a spectral-domain OCT scanner the spectral procedures equivalent to the above outlined formal Fourier transformations, an optical beam containing a set of $N$ discrete spectral components localized around a certain central frequency is radiated into the inspected tissue. Correspondingly, the wavenumbers of the probing-field spectral components are $k_{n} =k_{0} +\delta k\cdot n$, where the index $n$ ranges from $n=-(N-1)/2$ to $n=(N-1)/2$, so that the total spectrum width is $\Delta k_{n} =\delta k\cdot (N-1)$. To form an A-scan, the set of the backscattered optical spectral components is received such that both their amplitudes and phases are registered by the receiving sensor (or sensors) via comparison (heterodyning) with a reference beam. For the present consideration, technical details (e.g., using a swept source or a broadband source) of obtaining the necessary amplitudes and phases of $N$ spectral components are not essential.

To form an A-scan corresponding to an axially oriented array of spatial cells containing scatterers, the phase delays for all backscattered optical harmonics is natural to count from the surface of the tissue corresponding to the coordinate $z_{0} =0$. Next, in the ideal case it is assumed that OCT A-scans are formed by ballistic photons, i.e. photons that propagate in the axial direction of the beam and experience a single backscattering from the scatterers in the tissue bulk. The complex amplitude of each harmonic backscattered from a localized scatterer is determined by the scattering strength of the latter, as well as by the phase delay acquired during the optical wave propagation and determined by the depth of the scatterer. The reception of the backscattered beam is usually performed in the Fraunhofer zone, so that the phase difference related to the off-axial location of a scatterer inside the optical beam can be neglected (at least in the first approximation). Thus, the one-dimensional character of formation of axial A-scans is a good approximation to the reality. This fact ensures that properly performed summation of the received $N$ spectral harmonics backscattered from a localized obstacle located at $z=z_{0} $ can be considered as a fairly good approximation of the inverse 1D Fourier transform (IFT). The summation of such a discrete set of backscattered components correspondingly yields a discrete (pixelized) one-dimensional space-domain image (A-scan), in which the position of the scatterer located at $z=z_{0} $ is represented with an accuracy corresponding to the size of the spatial cells as depicted in Fig.~1. The scatterer localization in the transversal direction is determined by the lateral size of the optical beam.

The outlined above procedure of physical summation of spectral components is close, although not identical to formal inverse Fourier transform of the discrete spectrum of spatial harmonics corresponding to compact scatterers located along an 1D array of spatial cells. In particular, similarly to the formal forward and inverse Fourier transforms, the number $N$ of physically used spectral components determines the number of pixels (discrete spatial cells) in the constructed A-scan. However, there are also important differences between such a summation of physically backscattered waves and the formal inverse Fourier transform. Just these details are responsible for specific features of speckle patterns in real OCT images. 

The first key point is that the wavenumbers of real spectral components that form an OCT image are grouped around a certain central wavenumber $k_{0} =\omega _{0} /c$, where $\omega _{0} $ is the central optical frequency and $c$ is the phase velocity of light in the tissue. This is in contrast to the shown in Fig.~1 situation corresponding to the formal Fourier transforms, for which the wavenumbers are distributed around $k_{0} =\omega _{0} /c=0$. Certainly, for the recorded physical spectral components, fast oscillations of the received signal with the optical frequency $\omega _{0} $ are eliminated in the reception scheme by performing comparison with the phase of a reference optical beam using a heterodyning procedure. This procedure ensures that for $i$th scatterer located at the depth $z_{i} $, the acquired phase differences $(k_{n+1} -k_{n} )z_{i}^{} $ between all received spectral components are conserved similarly to the phase differences in the formal forward and inverse Fourier transforms depicted in Fig.~1. However, we recall that for the optical waves with the wavenumbers $k_{n} =k_{0} +\delta k_{n} $, the central wavenumber $k_{0} \ne 0$. Consequently, in the presence of several scatterers with coordinates $z_{i} \ne z_{j} $, after propagation forth and back, the corresponding backscattered waves acquire the phase delays $2k_{n} z_{i} =2(k_{0} +\delta k_{n} )z_{i} $ and $2k_{n} z_{j} =2(k_{0} +\delta k_{n} )z_{j} $. For the result of summation of such components, the contributions $2k_{0} z_{i} $ and $2k_{0} z_{j} $ related to the non-zero $k_{0} $ are important (for example, they may change the character of the backscattered wave interference if the inter-scatterer distance is less than the coherence length). If only a single scatterer is located at the depth $z=z_{0} $, this ``background'' phase delay $2k_{0} z_{0} $ is the same for all backscattered harmonics (a common phase factor) and consequently does not affect the amplitude (absolute value) of the reconstructed spatial A-scan representing such a single scatterer. However, if a group of several scatterers is located within the same spatial cell (sample volume) and the axial size $H/N$ of the spatial cells is not negligibly small compared with the optical wave length $\lambda _{0} =2\pi /k_{0} $ (which is a real situation in OCT), then the waves scattered by such closely located particles can interfere either destructively or constructively depending on the exact initial positions. Due to such interference the pixel may become either dark or bright. For example, the fields produced by these sub-resolution scatterers located within one sample volume may cancel each other in the initial state of the tissue, whereas in the deformed state may interfere in phase, producing peculiar blinking and ``boiling'' of speckles in the OCT image during the tissue deformation. Such blinking/boiling in some situations is considered as a kind of image-degrading noise or, in contrast, may be used as information-carrying effect. In particular, proper understanding of such phenomena is of primary interest for elastographic mapping in OCT, for which speckle decorrelation may be exploited as a useful effect [12]. 

Further, there is another important difference between the formal Fourier transforms and the physical summation of optical spectral components in OCT. Namely, for a particle located at $z=z_{0} $, the corresponding phase delay according to the Fourier-transform properties is $k_{n} z_{0}^{} $, whereas the physical backscattered waves received by the OCT scanner twice cover the distance $z=z_{0} $ (forth and back from the scatterer) and acquire the phase delays $2k_{n} z_{0}^{} $. Consequently, to ensure the correct spatial position of the scatterers in the reconstructed A-scan within the desired maximum depth range $0\le z\le H$, the double phase shift due to the wave propagation forth-and-back should be compensated by the twice smaller difference between the spectral components. Namely, the neighboring components should differ by the value $\delta k=k_{n+1} -k_{n} =\pi /H$ instead of $\delta k=k_{n+1} -k_{n} =2\pi /H$ in the formal Fourier transform depicted in Fig.~1. 

Taking all these features into account, for an A-scan formed via summation of real optical backscattered harmonics, instead of Eqs.\,((\ref{GrindEQ__1_})) and ((\ref{GrindEQ__2_})) for the formal Fourier transform one obtains 
\begin{equation} \label{GrindEQ__3_} 
\begin{array}{l} 
{A(q)=\sum _{j}\sum _{n}^{} A_{j} \exp (i2k_{n} z_{j} )\exp (-i{2\pi n} z_{q}/{H}) =} \\ {=\sum _{j}\sum _{n}^{} A_{j} \exp (i2k_{0} z_{j} )\exp (i{2\pi n} z_{j}/H )\exp (-i{2\pi n} z_{q}/H ) } 
\end{array} 
\end{equation} 
In Eq.~(\ref{GrindEQ__3_}) it is taken into account that $\delta k=k_{n+1} -k_{n} =\pi /H$ and $k_{n} =k_{0} +\pi n/H$. We underscore that in the right-hand side of Eq.~(\ref{GrindEQ__3_}), the positions $z_{j} $ of the scatterers correspond to their actual continuous coordinates in the physical space. In contrast, the coordinate $z_{q} $ for the pixelized amplitude function $A(q)$ in the constructed A-scan can take only a discrete number of values (corresponding to pixels in the OCT image). As has already been formulated above, the maximum number of these pixels is limited by the number $N$ of spectral optical components used in the OCT scanner. Therefore, the position of a small scatterer is known in the best case with an uncertainty equal to one pixel that corresponds to the size $H/(N-1)$ in the space domain. Despite such a discrete character of the constructed A-scan, even if the physical (continuous) coordinate $z_{j} $ of a localized scatterer changes to a value smaller than the axial size of the sample volume (coherence length), this small variation will manifest itself via the corresponding variation in the complex amplitude $A(q)$ due to the presence of the exponential factors $\exp (i2k_{0} z_{j} )\exp (i2\pi n z_{j}/H )$ in Eq.~(\ref{GrindEQ__3_}). This fact ensures the possibility to detect subpixel displacements of scatterers even in such pixelized images (even though the accuracy of determining the initial positions is limited by the pixel size).

 One should also take into account that the formal Fourier transform depicted in Fig.\,1 for a single scatterer, all spatial spectral harmonics have identical amplitudes (the envelope of the spectrum is rectangular). However, since the spectral form $S_{S} (k)$ of real optical sources is not necessarily rectangular, the corresponding amplitude factor should be introduced in Eq.(\ref{GrindEQ__3_}), so that the expression for an A-scan takes the form
\begin{equation} \label{GrindEQ__4_} 
A(q)=\sum _{j}\sum _{n}^{}S_{S} (k_{n} ) A_{j} \exp (i2k_{n} z_{j} )\exp (-i\frac{2\pi n}{H} z_{q} )  
\end{equation} 
Consequently, in contrast to the ideal rectangular spectral shape (that can ensure ultimately narrow localization within a single pixel number as is discussed above), a smoother spectrum $S_{S} (k_{n} )$ usually introduces somewhat stronger smoothing of a point-like particle image, so that speckle spots are spread between at least two adjacent pixels.

Expression (\ref{GrindEQ__4_}) for the A-scan readily allows one to study the influence of the source spectrum shape, as well as the roles of other key factors: the central wavelength, total spectral width (or equivalently, coherence length) of the sounding light, density of scatterers, number of spectral harmonics and pixels in the spatial-domain A-scan. Some other important characteristics can also be readily introduced. For example, the light polarization is not explicitly present in Eqs.~(\ref{GrindEQ__3_}) and (\ref{GrindEQ__4_}), which can be interpreted as the so-called co-polarization imaging in which scans are formed by the scattered light with the same linear polarization as in the incident wave. Certainly, in a similar manner, the scattering amplitudes $A_{j}^{co} $ and $A_{j}^{cross} $ into both co- and cross-polarized components can be readily attributed to the scatterers (such that the total scattering intensity is $A_{j}^{2} =(A_{j}^{co} )^{2} +(A_{j}^{cross} )^{2} $). Thus cross-polarized A-scans together with the scans based on the combining the cross- and co-polarized components can also be simulated. 

 Before turning to examples, we note that two-dimensional OCT images (B-scans) are composed as arrays of adjacent linear A-scans. Assuming different amplitude distribution in the transversal direction of the optical beam and different degree of lateral displacement between the adjacent A-scans one can readily simulate B-scans with more or less dense overlapping of A-scans. In the next section, for simplicity, we use the approximation of a uniform amplitude distribution in the cross section of the sounding beam and construct simulated B-scans by matching adjacent A-scans without overlapping (i.e. the scatterers within each A-scan are different). We show below that even this simple approximation well reproduces statistical properties of speckles in real OCT scans. \textbf{}

 \section{Examples of the model application for studying main characteristics of speckle patterns} 

Here, we present some examples illustrating the possibilities of the above-formulated model. Figure~2 demonstrates fragments (10x10 pixels to clearer show the differences) of simulated 2D speckle patterns corresponding to a random distribution of sub-resolution scatterers with different spatial densities. Each A-line contains $N=256$ pixels. To form an A-scan, a prescribed number $N_{sc} $ of scatterers with uniformly distributed random coordinates $z_{j} $ and equal scattering strength are numerically generated. The density of scatterers corresponding to Figs.~2 is $N_{sc} /N=$0.5, 1, 2 and 4.

\begin{figure}[htb]
\centerline{\includegraphics[width=13 cm]{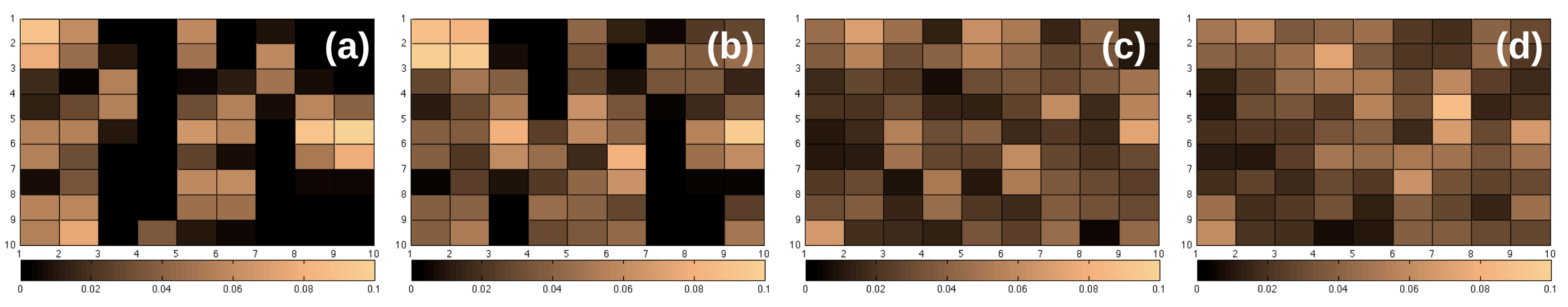}}
\begin{center}
\caption{Fragments (10x10 pixels) of simulated speckle patterns for $N=256$ pixels in an A-scan for randomly distributed $N_{sc} $ scatterers in each A-scan with the density $N_{sc} /N=$0.5, 1, 2 and 4 [plots (a), (b), (c) and (d), respectively]. The physical size of one pixel is 8 $\mu$m. The central optical wavelength in physical units is $\lambda _{0} =1$ $\mu$m and the coherence length $L_{c} $ is an order of magnitude greater than $\lambda _{0} $, which is a typical value for OCT scanners. To exclude the trivial increase of brightness with increasing density of scatterers, the amplitudes of the images are normalized (divided) by $(N_{sc} /N)^{^{1/2} } $.}
 \end{center}
\end{figure}
These patterns imitate scattering into the mode with the same polarization as in the incident wave. With increasing density $N_{sc} /N$ of scatterers the portion of dark (non-scattering) pixels obviously decreases. However, it can readily be verified that for the densities greater than $N_{sc} /N\sim 2$ per pixel, statistical properties of the speckle pattern weakly depend on the density of scatterers. This is shown in Fig.~3, where the amplitudes are normalized similarly to Fig.~2 and the total areas of the bars in the histograms are normalized to a unit value. For the very rarefied scatterers, the histograms demonstrate a significant percentage of nearly zero-amplitude pixels (in which the scatterers are absent). With increasing number of scatterers $N_{sc} /N\ge 2$, the shape of histograms stabilizes and, in agreement with the known speckle properties (see e.g. [3]), tends to a Rayleigh distribution (see the solid line in Fig.~3 shown in all panels for the same fitting parameters).

\begin{figure}[htb]
\centerline{\includegraphics[width=13 cm]{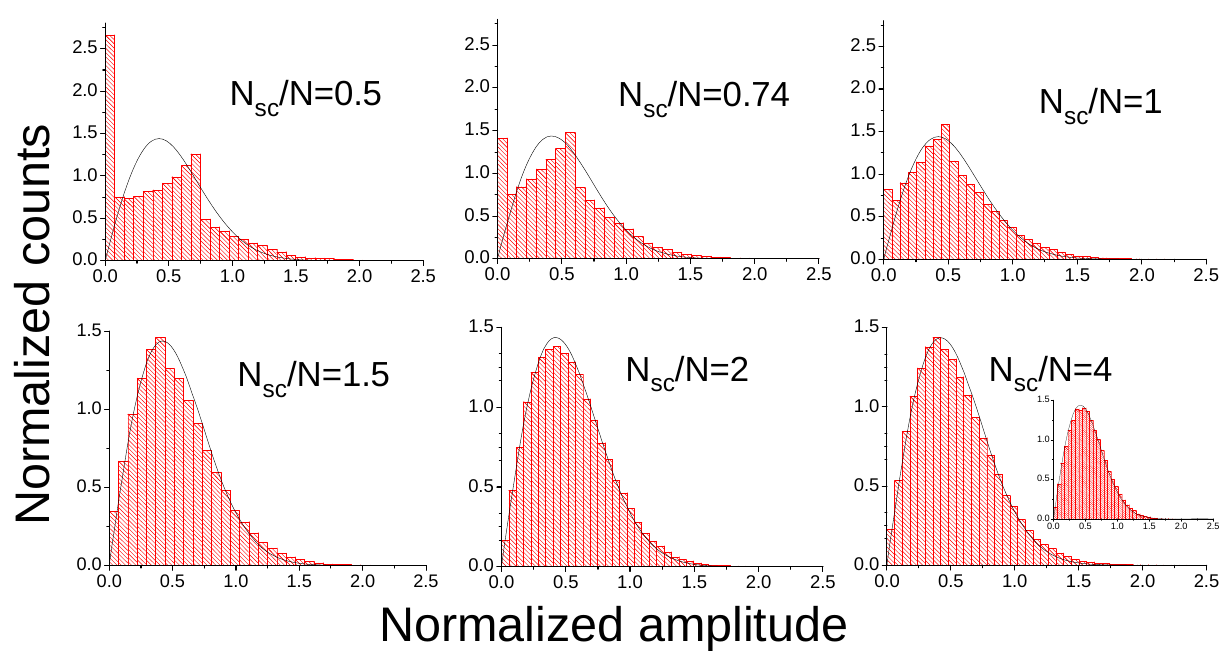}}
\begin{center}
\caption{
Simulated evolution of the pixel-amplitude histograms with increasing density of scatterers $N_{sc} /N=0.5;\; 0.74;\; 1.5;\; 2;\; 4\; \; $; and the bottom-right inset corresponds to the largest $N_{sc} /N=16$. For $N_{sc} /N\ge 1.5...2$ the histograms are well fitted by the Rayleigh distribution $(x/\sigma ^{2} )\exp (-x^{2} /2\sigma ^{2} )$ (solid line in all panels for the same fitting parameter $\sigma $). 
}
 \end{center}
\end{figure}
The next Fig.~4 demonstrates comparison with statistical properties of speckles in a real OCT scan obtained in the co-polarization mode). The simulated histograms in Fig.~4 are obtained for $N_{sc} /N=4$. Panels (a) and (c) are shown in the logarithmic scale to emphasize that the speckle intensities well obey exponential distributions [1]. 
\begin{figure}[htb]
\centerline{\includegraphics[width=9cm]{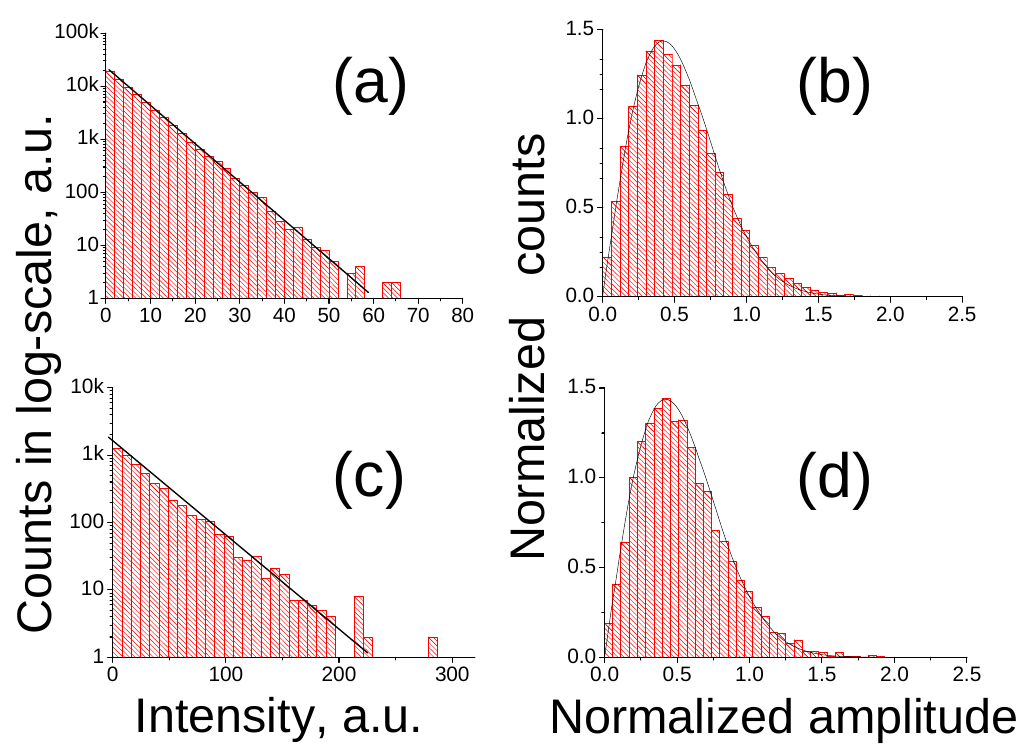}}
\caption{Histograms for the pixel intensities and amplitudes for simulated speckle patterns (upper raw) and an experimentally obtained OCT image (lower raw). Panels (a) and (c) are for the pixel intensities; panels (b) and (d) are for the pixel amplitudes. The dashed lines in the intensity histograms (a) and (c) are exponential fits and Rayleigh fits in the amplitude histograms (b) and (d).
}
\end{figure}

Further, Fig.~5 illustrates the simulation results for the influence of the ratio $L_{c} /\lambda _{0} $ between the coherence length $L_{c} $ and the central wavelength $\lambda _{0} $ (or equivalently for the varying ratio $k_{0} /\Delta k$). Figure 5 \begin{figure}[htb]
\centerline{\includegraphics[width=9cm]{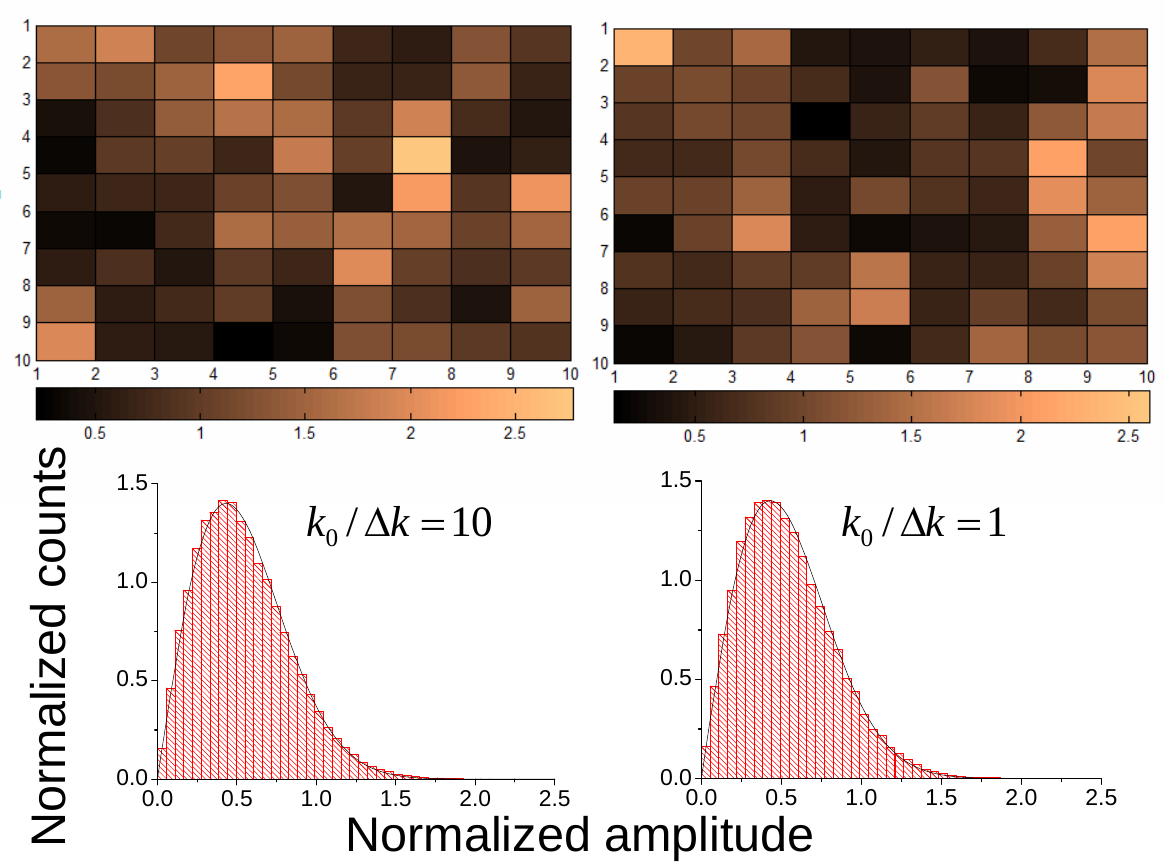}}
\caption{Fragments of simulated co-polarization images illustrating variability of the speckle-texture realizations for different ratios $k_{0} /\Delta k=10$ (left) and $k_{0} /\Delta k=1$ (right). In contrast, the respective speckle-amplitude histograms in both cases are well fitted by the same Rayleigh distribution. Simulations are made for $N_{sc} /N=4$.
}
\end{figure}presents the same fragment 10x10 pixels of the image simulated for the same positions of scatterers using an order of magnitude different ratios $k_{0} /\Delta k=10$ (typical of the majority of OCT systems) and $k_{0} /\Delta k=1$ (a super-broadband case). Even though the texture fragments look differently, this does not affect the statistical properties of the speckles as is seen from the shown below respective histograms of the pixel amplitudes. Both histograms are well fitted by a Rayleigh distribution with the same parameters. This is expectable, because the variation in the wavelength mainly changes the random phases of the backscattered waves, so that for sufficiently large number of such interfering waves, the average properties should not change.

In contrast to identical speckle statistics for static images, deformation of the material should produce strongly different evolution (blinking and boiling) of the speckle patterns for different ratios $k_{0} /\Delta k$ (or $L_{c} /\lambda _{0} $). Indeed, the same straining should cause very different variations in the mutual phasing of interfering waves scattered by sub-resolution scatterers located within the sample volume. Figure 6 shows the evolution of decorrelation between a reference OCT image and images of tissue subjected to straining in the vertical direction. 
\begin{figure}[htb]
\centerline{\includegraphics[width=12cm]{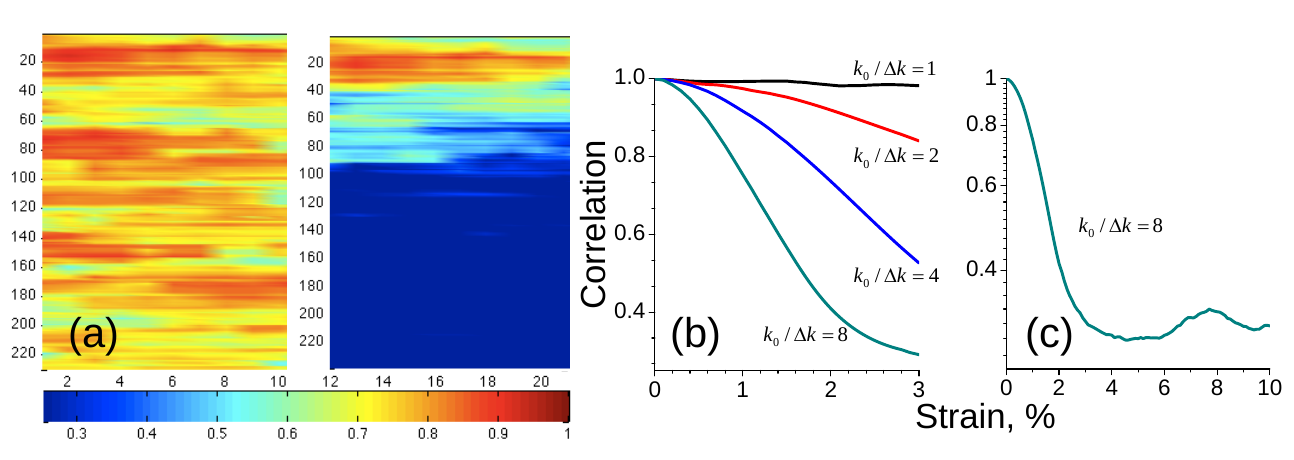}}
\caption{
Evolution of cross-correlation between a reference OCT image and those of a tissue subjected to uniaxial 1\% straining by the solid surface of OCT probe pressed from above: (a) - correlation maps after (left) and before (right) compensation of inhomogeneous strain-induced translational displacements; (b) - the cross-correlation coefficient averaged over the scan area (after compensation of translational displacements) as a function of uniaxial strain (up to 3\%) for different ratios $L_{c} /\lambda _{0} $ or equivalently different ratios $k_{0} /\Delta k$; (c) -- dependence similar to plot (b) for the ratio $k_{0} /\Delta k=10$ obtained in a wider strain range up to 10\% to illustrate the possibility of non-monotonic behavior of cross-correlation under monotonic strain increase. 
}
\end{figure}
Figure 6(a) shows the correlation maps for 1\% uniaxial straining with and without compensation the strain-induced inhomogeneous translational displacements. After performing the compensation the residual decorrelation depends only on the local strain and consequently can be used for plotting strain elastograms [12]. Figure 6(b) demonstrates that, as expected from simple analytical arguments \cite{ref24}, under the same strain, speckle blinking causing decorrelation is much more pronounced for larger ratios $L_{c} /\lambda _{0} $. The right scan demonstrates the effect of non-monotonic evolution of the cross-correlation caused by partial restoration of the speckle-spot intensities when the strain reaches a certain value $s_{ret} $ ensuring similar conditions of phasing between the scatterers located at the opposite sides of the sample volume, for which the phase difference between the scatterers changes by $2k_{0} L_{c} s_{ret} =2\pi $ rad. For approximately twice smaller strain $s_{\min } \approx s_{ret} /2$, the phasing changes by $\pi $ rad. and the character of interference changes to the opposite for a significant part of sampling volumes, so that the cross-correlation reaches minimum [see Fig.~6(b)].

 \section{Conclusions } 
Due to the used close-to-reality procedures, the proposed model generates images with the speckle structure as close as possible to images of real spectra-domain OCT scanners. Consequently the image processing principles tested on this model can be directly applied to real scans. An example of applying this model to the comparative analysis of elastographic mapping based on initial displacement-field reconstruction by correlation processing versus the direct use of the correlation-stability can be found in \cite{ref25}. Owing to similarity with formation of real scans, the used simulation procedures readily allowed us to study in detail how the deformation-induced speckle blinking and other measurement conditions (central optical frequency, coherence length, density of scatterers, displacement and strain values) affect the quality of displacement reconstruction by correlational speckle tracking.

In the above-considered variant the model reproduces the OCT image formation in the most pure form, i.e., assuming the ballistic character of optical-field propagation and single scattering from the localized scatterers into the same polarization. However, in a quite straightforward way the model allows for introducing additional factors important in real conditions, for example, additive noises of the receiving array, depth-dependence of the optical filed, scattering into the mode with the orthogonal polarization, additional random phase variation due to imperfectly straight trajectories of the photons or refraction index fluctuations. Taking into account heterogeneity of locations of scatterers and their distribution over scattering strengths is also straightforward. Introducing different profiles of the optical beam and different degree of overlapping between adjacent A-lines it is possible to imitate arbitrarily dense (rarefied) B-scans, beyond the assumption of a single form of point-spread function in the lateral direction. Due to its close-to-reality character, the proposed model can be useful for studying a wide range of problems (the above-mentioned elastographic mapping, testing of various methods of speckle suppression, studying correlation properties of scans and other aspects of OCT image processing). Certainly with appropriately adjusted parameters, among which the central wavelength and coherence length of the optical field are the key ones, this model can be readily applied to time-domain OCT images as well. 

 \ack 
The study was supported by the RFBR grants Nos 13-02-97131 and 13-02-00627 and the RF Government contract 14.B25.31.0015. V.Y.Z. and L.A.M. acknowledge the support of the grant of the RF Government No 11.G34.31.0066 and Russian Federation President grant for young researches MK-4826.2013.2, respectively.


\section*{References}
\begin{thebibliography}{99}

\bibitem{ref1}
Schmitt J M, Xiang S H and Yung K. M. 1999 \emph{J. Biomed. Opt.} \textbf{4} 95--105
\bibitem{ref2} 
Duncan D D and Kirkpatrick S J 2008 \emph{JOSA A } \textbf{25} 231-237
\bibitem{ref3}
Ramella-Roman Liu X, Huang J, Guo Y and Kang J 2013 \emph{JOSA A}  \textbf{30} 51-59  
\bibitem{ref4}  
Yao G. and Wang L. V. 1999 \emph{Physics in medicine and biology} \textbf{44} (9) 2307
\bibitem{ref5}  
Wang R. K. 2002 \emph{Physics in medicine and biology} \textbf{47}(13) 2281 

\bibitem{ref6} 
 
Pircher M., Go E., Leitgeb R., Fercher A. F. and Hitzenberger C. K. 2003 \emph{J. Biomed. Opt.} \textbf{8} (3) 565-569

\bibitem{ref7}  

Rong L., Xiao W., Pan F., Liu S. and Li R. 2010 \emph{Chinese Optics Letters} \textbf{8}(7) 653-655

\bibitem{ref8}  
Iftimia N., Tearney G. J. and Bouma B. E. 2003 \emph{J. Biomed. Opt.} \textbf{8}(2) 260-263

\bibitem{ref9} 
Ozcan A., Bilenca A., Desjardins A. E., Bouma B. E. and Tearney G. J. 2007 \emph{JOSA A} \textbf{24}(7) 1901-1910

\bibitem{ref10}  
Luan F. and Wu Y. 2013 \emph{Laser Phys. Lett.} \textbf{10}(3) 035603

\bibitem{ref11}
 
 Zaitsev V. Y., Matveev L. A., Gelikonov G. V, Matveyev A. L. and Gelikonov V. M. 2013 \emph{Laser Phys. Lett.} \textbf{10} 065601

\bibitem{ref12} 

Zaitsev V. Y., Matveev L. A., Matveyev A. L., Gelikonov G. V. and Gelikonov V. M. 2014 \emph{J. Biomed. Opt.} \textbf{19}(2) 21107

\bibitem{ref13} 
Fu J., Pierron F. and Ruiz P. D. 2013 \emph{J. Biomed. Opt.} \textbf{18}(12) 121512 

\bibitem{ref14}
Sun C., Standish B., Vuong B., Wen X. Y. and Yang, V. 2013 \emph{J. Biomed. Opt.} \textbf{18}(12) 121515

\bibitem{ref15}
Nahas A., Bauer M., Roux S. and Boccara A. C. 2013 \emph{Biomed. Opt. Express} \textbf{4}(10) 2138-2149 

\bibitem{ref16} 
Li J., Wang S., Singh M., Aglyamov S., Emelianov S., Twa M. D. and Larin K. V. 2014 \emph{Laser Phys. Lett.} \textbf{11}, 065601


\bibitem{ref17}

Kennedy B.F., Kennedy K.M., Sampson D.D. 2014 \emph{IEEE J. of Selected Topics in Quant. Electron.} \textbf{20}(2) 7101217

\bibitem{ref18}  
Wojtkowski M., Kowalczyk A., Leitgeb R. and Fercher A. F. 2002 \emph{Optics letters} \textbf{27}(16) 1415-1417

\bibitem{ref19}

Wojtkowski M., Leitgeb R., Kowalczyk A., Fercher A. F. and Bajraszewski T. 2002 \emph{J. Biomed. Opt.} \textbf{7}(3) 457-463

\bibitem{ref20} 
Leitgeb R., Hitzenberger C. and Fercher A. 2003 \emph{Optics Express} \textbf{11}(8) 889-894

\bibitem{ref21}
Yun S., Tearney G., de Boer J., Iftimia N. and Bouma B. 2003 \emph{Optics Express} \textbf{11}(22) 2953-2963
\bibitem{ref22}
G\"otzinger E., Pircher M., Leitgeb R. and Hitzenberger C. 2005 \emph{Optics express } \textbf{13}(2) 583-594
\bibitem{ref23}
Wang R. K. 2007 \emph{Applied Physics Letters} \textbf{90}(5) 054103

\bibitem{ref24}
Zaitsev V. Y. , Vitkin I.A. , Matveev L.A. , Gelikonov V.M. , Matveyev A.L.,  and Gelikonov G.V. 2014 \emph{Radiophys. and Quant. Electron.} \textbf{57}(3),  [in print]

\bibitem{ref25}
Zaitsev V.Y., Matveev L.A., Matveyev A.L., Gelikonov G.V. and Gelikonov V.M. 2014 \emph{Proc. SPIE 9129}  91290J

\end{thebibliography}
\end{document}